\documentclass{article}

\usepackage{bm}

\begin{document}

\title{Negative Energies in the Dirac equation}         
\author{Valeriy V. Dvoeglazov\\
Universidad de Zacatecas\\
Apartado Postal 636, Suc. 3\\
Zacatecas 98061, Zac., M\'exico\\
E-mail: valeri@fisica.uaz.edu.mx}        
\date{\empty}          
\maketitle

\begin{abstract}
It is easy to check that both algebraic equation  $Det (\hat p - m) =0$ and $Det (\hat p + m) =0$  for  $u-$ and $v-$  4-spinors have solutions with $p_0= \pm E_p =\pm \sqrt{{\bf p}^2 +m^2}$. The same is true for higher-spin equations. Meanwhile, every book considers
 the equality $p_0=E_p$  for both $u-$  and $v-$ spinors of the $(1/2,0)\oplus (0,1/2))$ representation only, thus applying the Dirac-Feynman-Stueckelberg procedure for elimination of the negative-energy solutions. The recent Ziino works (and, independently, 
the articles of several others) show that the Fock space can be doubled. We re-consider this possibility on the quantum field level for both  $s=1/2$ and higher spin particles.
\end{abstract}


\large{

\section{Introduction}

The recent problems of superluminal neutrinos, e.~g., Ref.~\cite{n1}, negative-mass squared neutrinos, e.~g.~\cite{n2}, various schemes of oscillations including sterile neutrinos, e.~g.~\cite{n3}, require much attention.
The problem of the lepton mass splitting ($e, \mu, \tau$) has long history~\cite{ms}. This suggests that something missed in the foundations of relativistic quantum theories. Modifications seem to be necessary in the Dirac sea concept, and in the even more sophisticated 
Stueckelberg concept of the backward propagation in time. The Dirac sea concept is intrinsically related to the Pauli principle. However, the Pauli principle is intrinsically connected with the Fermi statistics and the anticommutation relations of fermions. Recently, the concept of the {\it bi-orthonormality} has been proposed; the (anti) commutation relations and statistics are assumed to be different for {\it neutral} particles~\cite{Ahlnew}.
One can speculate that they go off in the negative-energy sea, but due to some reasons (interaction?) they do not live 
there (from our viewpoint), but return back
(been expelled), thus showing us the new kind of oscillations on the Planck scale $\omega >> E/\hbar$, Ref.~\cite{Dvonew}.
Perhaps, some of the neutrinos remain {\it sterile} even in our world.

We propose the relevant modifications in the basics of the relativistic quantum theory below. However much work is still needed.

\section{The General Framework and Connections with Previous Models}

The Dirac equation is:
\begin{equation}
[i\gamma^\mu \partial_\mu -m]\Psi (x) =0\,.\label{Dirac}
\end{equation}
At least, 3 methods of its derivation exist~\cite{Dirac,Sakurai,Ryder}:
\begin{itemize}
  \item the Dirac one (the Hamiltonian should be linear in $\partial/\partial x^i$, and be compatible with $E_p^2 -{\bf p}^2 c^2 =m^2 c^4$);
  \item the Sakurai one (based on the equation $(E_p- {\bf \sigma} \cdot {\bf p}) (E_p+ {\bf \sigma} \cdot {\bf p}) \phi =m^2 \phi$);
  \item the Ryder one (the relation between  2-spinors at rest is $\phi_R ({\bf 0}) = \pm \phi_L ({\bf 0})$ and boosts).
\end{itemize}
The $\gamma^\mu$ are the Clifford algebra matrices 
\begin{equation}
\gamma^\mu \gamma^\nu +\gamma^\nu \gamma^\mu = 2g^{\mu\nu}\,.
\end{equation}
Usually, everybody uses the following definition of the field operator~\cite{Itzykson} in the pseudo-Euclidean metrics:
\begin{equation}
\Psi (x) = \frac{1}{(2\pi)^3}\sum_h \int \frac{d^3 {\bf p}}{2E_p} [ u_h ({\bf p}) a_h ({\bf p}) e^{-ip\cdot x}
+ v_h ({\bf p}) b_h^\dagger ({\bf p})] e^{+ip\cdot x}]\,,
\end{equation}
as given {\it ab initio}.
After actions of the Dirac operator at  $\exp (\mp ip_\mu x^\mu)$ the 4-spinors ( $u-$ and $v-$ ) 
satisfy the momentum-space equations: $(\hat p - m) u_h (p) =0$ and \linebreak $(\hat p + m) v_h (p) =0$, respectively; the $h$ is 
the polarization index. It is easy to prove from the characteristic equations
$Det (\hat p \mp m) =(p_0^2 -{\bf p}^2 -m^2)^2= 0$ that the solutions should satisfy the energy-momentum relation $p_0= \pm E_p =\pm \sqrt{{\bf p}^2 +m^2}$.

The general scheme of construction of the field operator has been presented in~\cite{Bogoliubov}. In the case of
the $(1/2,0)\oplus (0,1/2)$ representation we have:
\begin{eqnarray}
&&\Psi (x) = {1\over (2\pi)^3} \int d^4 p \,\delta (p^2 -m^2) e^{-ip\cdot x}
\Psi (p) =\nonumber\\
&=& {1\over (2\pi)^3} \sum_{h}^{}\int d^4 p \, \delta (p_0^2 -E_p^2) e^{-ip\cdot x}
u_h (p_0, {\bf p}) a_h (p_0, {\bf p}) =\label{fo}\\
&=&{1\over (2\pi)^3} \int {d^4 p \over 2E_p} [\delta (p_0 -E_p) +\delta (p_0 +E_p) ] 
[\theta (p_0) +\theta (-p_0) ] e^{-ip\cdot x}
\sum_{h}^{} u_h (p) a_h (p) \nonumber\\
&=& {1\over (2\pi)^3} \sum_h^{} \int {d^4 p \over 2E_p} [\delta (p_0 -E_p) +\delta (p_0 +E_p) ] 
\left
[\theta (p_0) u_h (p) a_h (p) e^{-ip\cdot x}  + \right.\nonumber\\
&+&\left.\theta (p_0) u_h (-p) a_h (-p) e^{+ip\cdot x} \right ] 
= {1\over (2\pi)^3} \sum_h^{} \int {d^3 {\bf p} \over 2E_p} \theta(p_0)  
\left [ u_h (p) a_h (p)\vert_{p_0=E_p} e^{-i(E_p t-{\bf p}\cdot {\bf x})}  +\right.\nonumber\\ 
&+& \left. u_h (-p) a_h (-p)\vert_{p_0=E_p} e^{+i (E_p t- {\bf p}\cdot {\bf x})} 
\right ]\nonumber
\end{eqnarray}
During the calculations above we had to represent $1=\theta (p_0) +\theta (-p_0)$
in order to get positive- and negative-frequency parts.\footnote{See Ref.~\cite{DvoeglazovJPCS}
for some discussion.} Moreover, during these calculations we did not yet assumed, which equation this
field operator  (namely, the $u-$ spinor) satisfies, with negative- or positive- mass? 

In general we should transform $u_h (-p)$ to the $v (p)$. The procedure is the following one~\cite{DvoeglazovHJ}.
In the Dirac case we should assume the following relation in the field operator:
\begin{equation}
\sum_{h}^{} v_h (p) b_h^\dagger (p) = \sum_{h}^{} u_h (-p) a_h (-p)\,.\label{dcop}
\end{equation}
We know that~\cite{Ryder}
\begin{eqnarray}
\bar u_\mu (p) u_\lambda (p) &=& +m \delta_{\mu\lambda}\,,\\
\bar u_\mu (p) u_\lambda (-p) &=& 0\,,\\
\bar v_\mu (p) v_\lambda (p) &=& -m \delta_{\mu\lambda}\,,\\
\bar v_\mu (p) u_\lambda (p) &=& 0\,,
\end{eqnarray}
but we need $\Lambda_{\mu\lambda} (p) = \bar v_\mu (p) u_\lambda (-p)$.
By direct calculations,  we find
\begin{equation}
-mb_\mu^\dagger (p) = \sum_{\lambda}^{} \Lambda_{\mu\lambda} (p) a_\lambda (-p)\,.
\end{equation}
Hence, $\Lambda_{\mu\lambda} = -im ({\bm \sigma}\cdot {\bf n})_{\mu\lambda}$, ${\bf n} = {\bf p}/\vert{\bf p}\vert$, 
and 
\begin{equation}
b_\mu^\dagger (p) = i\sum_\lambda ({\bm\sigma}\cdot {\bf n})_{\mu\lambda} a_\lambda (-p)\,.
\end{equation}
Multiplying (\ref{dcop}) by $\bar u_\mu (-p)$ we obtain
\begin{equation}
a_\mu (-p) = -i \sum_{\lambda} ({\bm \sigma} \cdot {\bf n})_{\mu\lambda} b_\lambda^\dagger (p)\,.
\end{equation}
The equations are self-consistent.\footnote{In the $(1,0)\oplus (0,1)$ representation 
the similar procedure leads to somewhat different situation:
\begin{equation}
a_\mu (p) = [1-2({\bf S}\cdot {\bf n})^2]_{\mu\lambda} a_\lambda (-p)\,. 
\end{equation}
This signifies that in order to construct the Sankaranarayanan-Good field operator (which was used by Ahluwalia, Johnson and Goldman [Phys. Lett. B (1993)], it satisfies 
$[\gamma_{\mu\nu} \partial_\mu \partial_\nu - {(i\partial/\partial t)\over E} 
m^2 ] \Psi (x) =0$, we need additional postulates. For instance, one can try to construct 
the left- and the right-hand side of the field operator separately each other~\cite{DvoeglazovJPCS}.}

However, other ways of thinking are possible. First of all to mention, we have, in fact,
$u_h ( E_p, {\bf p} )$ and $u_h (-E_p, {\bf p})$ originally, 
which may satisfy the equations:\footnote{Remember that, as before, we can always make the substitution 
${\bf p}\rightarrow -{\bf p}$ in any of the integrands of (\ref{fo}).}
\begin{equation}
\left [ E_p (\pm \gamma^0) - {\bm \gamma}\cdot {\bf p} - m \right ] u_h (\pm E_p, {\bf p})=0\,.
\end{equation}
Due to the properties $U^\dagger \gamma^0 U=-\gamma^0$, $U^\dagger \gamma^i U=+\gamma^i$
with the unitary matrix $U=\pmatrix{0&-1\cr 1&0\cr}= \gamma^0\gamma^5$ in 
the Weyl basis,\footnote{The properties of the $U-$ matrix are opposite to those of
$P^\dagger \gamma^0 P=+\gamma^0$, $P^\dagger \gamma^i P=-\gamma^i$
with the usual $P=\gamma^0$, thus giving $\left [ -E_p \gamma^0 + {\bf \gamma}\cdot {\bf p} - m \right ] 
P u_h (- E_p, {\bf p}) = -\left [\hat p +m \right ] \tilde v_{?} (E_p, {\bf p}) = 0$. While, the relations of the spinors $v_h (E_p, {\bf p})=\gamma_5  u_h (E_p, {\bf p})$ are well-known, it seems that
the relations of the $v-$ spinors of the positive energy to $u-$ spinors of the negative energy
are frequently forgotten, $\tilde v_{?} (E_p, {\bf p}) = \gamma^0 u_h (- E_p, {\bf p})$. } 
we have
\begin{equation}
\left [ E_p \gamma^0 - {\bm \gamma}\cdot {\bf p} - m \right ] U^\dagger u_h (- E_p, {\bf p})=0\,.\label{nede}
\end{equation}
Thus, unless the unitary transformations do not change
the physical content, we have that the negative-energy spinors $\gamma^5 \gamma^0 u^-$ (see (\ref{nede})) satisfy the accustomed ``positive-energy" Dirac equation. Their explicite forms $\gamma^5 \gamma^0 u^-$ are different from the textbook ``positive-energy" Dirac spinors.
They are the following ones:\footnote{We use tildes because we do not 
yet know their polarization properties.}
\begin{eqnarray}
\tilde u (p) = \frac{N}{\sqrt{2m (-E_p +m)}} \pmatrix{-p^+ + m\cr -p_r\cr
p^- -m \cr - p_r\cr}\,,\\
\tilde{\tilde u} (p) =\frac{N}{\sqrt{2m (-E_p +m)}}\pmatrix{-p_l \cr -p^- + m\cr
-p_l \cr p^+ -m\cr}\,.
\end{eqnarray}
$E_p=\sqrt{{\bf p}^2 +m^2}>0$, $p_0=\pm E_p$, $p^\pm = E\pm p_z$, $p_{r,l}= p_x\pm ip_y$.
Their normalization is to $(-2N^2)$.

What about the $\tilde v (p)=\gamma^0 u^-$ transformed with the $\gamma_0$ matrix?
Are they equal to $ v_h (p) =\gamma^5 u_h (p)$? Our answer is `{\it no}'.
Obviously, they also do not have well-known forms  of the usual $v-$ spinors in the Weyl basis, 
differing by phase factor and in the sign at the mass term (!)

Next, one can prove that the matrix
\begin{equation}
P= e^{i\theta}\gamma^0 = e^{i\theta}\pmatrix{0& 1_{2\times 2}\cr 1_{2\times 2} & 0\cr}
\label{par}
\end{equation}
can be used in the parity operator as well as
in the original Weyl basis. The parity-transformed function
$\Psi^\prime (t, -{\bf x})=P\Psi (t,{\bf x})$ must satisfy
\begin{equation}
[i\gamma^\mu \partial_\mu^{\,\prime} -m ] \Psi^\prime (t,-{\bf x}) =0 \,,
\end{equation}
with $\partial_\mu^{\,\prime} = (\partial/\partial t, -{\bf \nabla}_i)$.
This is possible when $P^{-1}\gamma^0 P = \gamma^0$ and
$P^{-1} \gamma^i P = -\gamma^i$. The matrix (\ref{par})
satisfies these requirements, as in the textbook case.
However, if we would take the phase factor to be zero
we obtain that while $u_h (p)$ have the eigenvalue $+1$, but ($R= ({\bf x} \rightarrow -{\bf x}, {\bf p} 
\rightarrow -{\bf p}$))
\begin{equation}
P R\tilde u (p) = P R\gamma^5 \gamma^0  u (-E_p, {\bf p})= -\tilde u (p)\,,\quad 
P R \tilde{\tilde u} (p) = P R \gamma^5 \gamma^0  u (-E_p, {\bf p})= -\tilde{\tilde u} (p)\,.
\end{equation}
Perhaps, one should choose the phase factor $\theta=\pi$. Thus, we again confirmed
that the relative (particle-antiparticle) intrinsic parity has physical significance only.

Similar formulations have been  presented in Refs.~\cite{Markov}, 
and~\cite{BarutZiino}. The group-theoretical basis for such doubling has been given
in the papers by Gelfand, Tsetlin and Sokolik~\cite{Gelfand}, who first presented 
the theory in the 2-dimensional representation of the inversion group in 1956 (later called as ``the Bargmann-Wightman-Wigner-type quantum field theory" in 1993). M. Markov wrote long ago {\it two} Dirac equations with  the opposite signs at the mass term~\cite{Markov}. 
\begin{eqnarray}
\left [ i\gamma^\mu \partial_\mu - m \right ]\Psi_1 (x) &=& 0\,,\\
\left [ i\gamma^\mu \partial_\mu + m \right ]\Psi_2 (x) &=& 0\,.
\end{eqnarray}
In fact, he studied all properties of this relativistic quantum model (while he did not know yet the quantum
field theory in 1937). Next, he added and  subtracted these equations. What did he obtain?
\begin{eqnarray}
i\gamma^\mu \partial_\mu \varphi (x) - m \chi (x) &=& 0\,,\\
i\gamma^\mu \partial_\mu \chi (x) - m \varphi (x) &=& 0\,.
\end{eqnarray}
Thus, $\varphi$ and $\chi$ solutions can be presented as some superpositions of the Dirac 4-spinors $u-$ and $v-$.
These equations, of course, can be identified with the equations for the Majorana-like $\lambda -$ and $\rho -$, which we presented 
in Ref.~\cite{DvoeglazovNP}.\footnote{Of course, the signs at the mass terms
depend on, how do we associate the positive- or negative- frequency solutions with $\lambda$ and $\rho$.}
\begin{eqnarray}
i \gamma^\mu \partial_\mu \lambda^S (x) - m \rho^A (x) &=& 0 \,,
\label{11}\\
i \gamma^\mu \partial_\mu \rho^A (x) - m \lambda^S (x) &=& 0 \,,
\label{12}\\
i \gamma^\mu \partial_\mu \lambda^A (x) + m \rho^S (x) &=& 0\,,
\label{13}\\
i \gamma^\mu \partial_\mu \rho^S (x) + m \lambda^A (x) &=& 0\,.
\label{14}
\end{eqnarray}
Neither of them can be regarded as the Dirac equation.
However, they can be written in the 8-component form as follows:
\begin{eqnarray}
\left [i \Gamma^\mu \partial_\mu - m\right ] \Psi_{_{(+)}} (x) &=& 0\,,
\label{psi1}\\
\left [i \Gamma^\mu \partial_\mu + m\right ] \Psi_{_{(-)}} (x) &=& 0\,,
\label{psi2}
\end{eqnarray}
with
\begin{eqnarray}
&&\hspace{-20mm}\Psi_{(+)} (x) = \pmatrix{\rho^A (x)\cr
\lambda^S (x)\cr},
\Psi_{(-)} (x) = \pmatrix{\rho^S (x)\cr
\lambda^A (x)\cr}, \quad\mbox{and}\,\Gamma^\mu =\pmatrix{0 & \gamma^\mu\cr
\gamma^\mu & 0\cr}\,.
\end{eqnarray}
It is easy to find the corresponding projection operators, and the Feynman-Stueckelberg propagator.

You may say that all this is just related to the spin-parity basis rotation (unitary transformations). However, 
in the previous papers I explained: the connection with the Dirac spinors has 
been found~\cite{DvoeglazovNP,Kirchbach}.\footnote{I also acknowledge
personal communications from D. V. Ahluwalia on these matters.}
For instance,
\begin{eqnarray}
\pmatrix{\lambda^S_\uparrow ({\bf p}) \cr \lambda^S_\downarrow ({\bf p}) \cr
\lambda^A_\uparrow ({\bf p}) \cr \lambda^A_\downarrow ({\bf p})\cr} = {1\over
2} \pmatrix{1 & i & -1 & i\cr -i & 1 & -i & -1\cr 1 & -i & -1 & -i\cr i&
1& i& -1\cr} \pmatrix{u_{+1/2} ({\bf p}) \cr u_{-1/2} ({\bf p}) \cr
v_{+1/2} ({\bf p}) \cr v_{-1/2} ({\bf p})\cr},\label{connect}
\end{eqnarray}
provided that the 4-spinors have the same physical dimension.
Thus, we can see
that the two 4-spinor systems are connected by the unitary transformations, and this represents
itself the rotation of the spin-parity basis. However, it is usually assumed that the $\lambda-$ and $\rho-$ spinors describe the neutral particles,
meanwhile $u-$ and $v-$ spinors describe the charged particles. Kirchbach~\cite{Kirchbach} found the amplitudes for 
neutrinoless double beta decay ($00\nu\beta$) in this scheme. It is obvious from (\ref{connect}) that there are some additional terms comparing with the standard formulation.  

One can also re-write the above equations into the two-component forms. Thus, one obtains the Feynman-Gell-Mann~\cite{FG} 
equations.
As Markov wrote himself, he was expecting ``new physics" from these equations. 

Barut and Ziino~\cite{BarutZiino} proposed yet another model. They considered
$\gamma^5$ operator as the operator of the charge conjugation. Thus, the charge-conjugated
Dirac equation has the different sign comparing with the ordinary formulation:
\begin{equation}
[i\gamma^\mu \partial_\mu + m] \Psi_{BZ}^c =0\,,
\end{equation}
and the so-defined charge conjugation applies to the whole system, fermion+electro\-magnetic field, $e\rightarrow -e$
in the covariant derivative. The superpositions of the $\Psi_{BZ}$ and $\Psi_{BZ}^c$ also give us 
the ``doubled Dirac equation", as the equations for $\lambda-$ and $\rho-$ spinors. 
The concept of the doubling of the Fock space has been
developed in the Ziino works (cf.~\cite{Gelfand,DvoeglazovBW}) in the framework of the quantum field theory. In their case the charge conjugate states
are simultaneously the eigenstates of the chirality.
Next, it is interesting to note that for the Majorana-like field operators we have
\begin{eqnarray}
\lefteqn{\hspace{-25mm}\left [ \nu^{^{ML}} (x^\mu) + {\cal C} \nu^{^{ML\,\dagger}} (x^\mu) \right
]/2 = \int {d^3 {\bf p} \over (2\pi)^3 } {1\over 2E_p} \sum_\eta \left
[\pmatrix{i\Theta \phi_{_L}^{\ast \, \eta} (p^\mu) \cr 0\cr} a_\eta
(p^\mu)  e^{-ip\cdot x} +\right.}\nonumber \\
&&+\left.\pmatrix{0\cr
\phi_L^\eta (p^\mu)\cr } a_\eta^\dagger (p^\mu) e^{ip\cdot x} \right ]\,
,\\
\lefteqn{\hspace{-25mm}\left [\nu^{^{ML}} (x^\mu) - {\cal C} \nu^{^{ML\,\dagger}} (x^\mu) \right
]/2 = \int {d^3 {\bf p} \over (2\pi)^3 } {1\over 2E_p} \sum_\eta \left
[\pmatrix{0\cr \phi_{_L}^\eta (p^\mu) \cr } a_\eta (p^\mu)  e^{-ip\cdot x}
+\right.}\nonumber\\
&&+\left.\pmatrix{-i\Theta \phi_{_L}^{\ast\, \eta} (p^\mu)\cr 0
\cr } a_\eta^\dagger (p^\mu) e^{ip\cdot x} \right ]\, , 
\end{eqnarray}
which, thus, naturally lead to the Ziino-Barut scheme of massive chiral
fields, Ref.~\cite{BarutZiino}.

Finally, I would like to mention that, in general,  in the Weyl basis the $\gamma-$ matrices are  {\it not} Hermitian, 
$\gamma^{\mu^\dagger}  =\gamma^0 \gamma^\mu \gamma^0$. So, $\gamma^{i^\dagger}= -\gamma^i$, $i=1,2,3$, the pseudo-Hermitian matrix.
The energy-momentum operator $i\partial_\mu$ is obviously Hermitian. So, the question, if the eigenvalues of 
the Dirac operator
(the mass, in fact) would be always real? The question of the complete system of the eigenvectors of the 
{\it non-}Hermitian operator deserve careful consideration~\cite{Ilyin}.  As mentioned before, 
Bogoliubov and Shirkov~\cite[p.55-56]{Bogoliubov} used the scheme to construct the complete set
of solutions of the relativistic equations, fixing the sign of $p_0=+E_p$. 

\section{Conclusions}       

The main points of my paper are: there are ``negative-energy solutions" in that is previously  considered as 
``positive-energy solutions" of relativistic wave equations, and vice versa. Their explicit forms have been presented
in the case of spin-1/2. 
Next, the relations to the previous works have been found. For instance,
the doubling of the Fock space and the corresponding solutions of the Dirac equation
obtained  additional mathematical bases in this paper. Similar conclusion can be deduced for
the higher-spin equations. 

I appreciate the discussions with participants of several recent Conferences.
}


\begin{thebibliography}{99}

\bibitem{n1} Guang-jiong Ni,  In {\it Relativity, Gravitation, Cosmology.} (Nova Science Pub., Huntington, NY, USA, 2004), hep-ph/0306028.

\bibitem{n2} N. Simicevic, {\it Implications of Weak-Interaction Space Deformation for Neutrino Mass
Measurements}, nucl-th/9909034.

\bibitem{n3} S. M. Bilenky and C. Giunti, {\it Nucl. Phys. Proc. Suppl.} {\bf  48}, 381 (1996).

\bibitem{ms} A. O. Barut, {\it Phys. Lett.} B{\bf 73}, 310  (1978); {\it Phys. Rev. Lett.} {\bf 42}, 1251 (1979).

\bibitem{Ahlnew} D. V. Ahluwalia and A. Ch. Nayak,  Int. J. Mod. Phys. D{\bf  23}, 1430026 (2014).

\bibitem{Dvonew} V. V. Dvoeglazov, Adv. Appl. Clifford Algebras (2016), accepted.

\bibitem{Dirac} P. A. M. Dirac, {\it Proc. Roy. Soc. Lond.} A{\bf 117}, 610 (1928). 

\bibitem{Sakurai}  J. J. Sakurai, {\it Advanced Quantum Mechanics.} (Addison-Wesley, 1967).

\bibitem{Ryder} L. H. Ryder, {\it Quantum Field Theory.} (Cambridge University Press, Cambridge, 1985).

\bibitem{Itzykson} C. Itzykson and J.-B. Zuber, {\it Quantum Field Theory.} (McGraw-Hill Book Co., 1980), p. 156.

\bibitem{Bogoliubov} N. N. Bogoliubov and D. V. Shirkov, {\it Introduction to the Theory of Quantized Fields.} 2nd Edition. (Nauka, Moscow, 1973).

\bibitem{DvoeglazovJPCS} V. V. Dvoeglazov, {\it J. Phys. Conf. Ser.} {\bf 284}, 012024  (2011),  arXiv:1008.2242.

\bibitem{DvoeglazovHJ} V. V. Dvoeglazov, {\it Hadronic J. Suppl.} {\bf 18}, 239 (2003), physics/0402094; {\it Int. J. Mod. Phys.} 
B{\bf 20}, 1317 (2006).

\bibitem{Markov} M. Markov, {\it ZhETF} {\bf 7}, 579 (1937); ibid. 603; {\it Nucl. Phys.} {\bf 55}, 130 (1964).
 
\bibitem{BarutZiino} A. Barut and G. Ziino, {\it Mod. Phys. Lett.} A{\bf 8}, 1099 (1993); G. Ziino, {\it Int. J. Mod. Phys.}
A {\bf 11},  2081 (1996).

\bibitem{Gelfand} I. M. Gelfand and M. L. Tsetlin, {\it ZhETF} {\bf 31} (1956) 1107; G. A. Sokolik, {\it ZhETF} {\bf 33} (1957) 1515.

\bibitem{DvoeglazovNP} V. V. Dvoeglazov, {\it Int. J. Theor. Phys.} {\bf 34} (1995) 2467; {\it Nuovo Cim.} {\bf 108}A (1995) 1467;  {\it Hadronic J.} {\bf 20} (1997) 435; {\it Acta Phys. Polon.} B{\bf 29} (1998) 619.

\bibitem{FG} R. P. Feynman and M. Gell-Mann, {\it Phys. Rev.} {\bf 109} (1958) 193.

\bibitem{Kirchbach} M. Kirchbach, C. Compean and L. Noriega, {\it Eur. Phys. J.} A{\bf 22} (2004) 149.

\bibitem{DvoeglazovBW} V. V. Dvoeglazov, {\it Int. J. Theor. Phys.} {\bf 37} (1998) 1915.

\bibitem{Ilyin} V. A. Ilyin, {\it Spektralnaya Teoriya Differencialnyh Operatorov.} (Nauka, Moscow, 1991); 
V. D. Budaev, {\it Osnovy Teorii Nesamosopryazhennyh Differencialnyh Operatorov.} (SGMA, Smolensk, 1997).

\end{thebibliography}
\end{document}